\documentclass[]{interact}

\usepackage{epstopdf}
\usepackage[caption=false]{subfig}

\usepackage[longnamesfirst,sort]{natbib}
\bibpunct[, ]{(}{)}{;}{a}{,}{,}

\theoremstyle{plain}

\theoremstyle{definition}

\theoremstyle{remark}

\usepackage[hidelinks]{hyperref}
\usepackage{natbib}

\DeclareMathAlphabet{\mathpzc}{OT1}{pzc}{m}{it}

\newcommand{\targ}{{\mathsf{targ}}}
\newcommand{\axis}{{\mathfrak{a}}}
\newcommand{\vol}{{\mathsf{vol}}}
\newcommand{\map}{{\mathpzc{T}}}
\newcommand{\init}{{\mathsf{init}}}
\newcommand{\grip}{{\mathsf{grip}}}
\newcommand{\fix}{{\mathsf{fix}}}

\newcommand{\conf}{{\mathsf{C}}}
\newcommand{\obs}{{\mathcal{O}}}
\newcommand{\free}{{\mathcal{F}}}
\newcommand{\cost}{{\mathsf{cost}}}

\newcommand{\com}[1]{}

\begin{document}


\title{Automated Process Planning for Turning: A Feature-Free Approach}

\author{Author: Anonymous for blind review}

\author{ \name{Morad Behandish\textsuperscript{a}\thanks{CONTACT Morad Behandish. Email: moradbeh@parc.com}, Saigopal Nelaturi\textsuperscript{a}, Chaman Singh Verma\textsuperscript{a}, and Mats Allard\textsuperscript{b}} \affil{\textsuperscript{a}Palo Alto Research Center, Palo Alto, CA, USA; \textsuperscript{b}Sandvik Coromant, Sandviken, Sweden}}

\maketitle

\begin{abstract}
	Turning is the most commonly available and least expensive machining operation, in terms of both machine-hour rates and tool insert prices. A practical CNC process planner has to maximize the utilization of turning, not only to attain precision requirements for turnable surfaces, but also to minimize the machining cost, while non-turnable features can be left for other processes such as milling. Most existing methods rely on separation of surface features and lack guarantees when analyzing complex parts with interacting features. In a previous study, we demonstrated successful implementation of a feature-free milling process planner based on configuration space methods used for spatial reasoning and AI search for planning. This paper extends the feature-free method to include turning process planning. It opens up the opportunity for seamless integration of turning actions into a mill-turn process planner that can handle arbitrarily complex shapes with or without a priori knowledge of feature semantics.
\end{abstract}

\begin{keywords}
	Computer-Aided Process Planning; Machining; Turning; Configuration Space
\end{keywords}

\section {Introduction}

The manufacturing technology has diversified dramatically over the past several decades with the increasing popularity of additive processes. Nevertheless, subtractive processes (i.e., machining) remain central to many industrial-scale, mission-critical, and high-precision mechanical parts.

Computer-aided process planning (CAPP) is the systematic determination of a set of steps by which a product can be manufactured in a cost-effective, competitive manner \citep{Alting1989computer}.
The most common approaches to CAPP are based on automated feature recognition \citep{Weill1982survey,Shah1991survey,Babic2008review,Xu2011computer}. These methods include convex volume decomposition \citep{Perng1990automatic,Kim1992recognition,Eftekharian2012convex}, graph-based heuristics \citep{Joshi1988graph,Wu1996analysis,Fu2012graph}, rule-based pattern recognition \citep{Vandenbrande1993spatial,Babic2008review}, among others. Their main challenge is that the notion of a ``feature'' is typically defined in an application-specific context \citep{Hoffmann2016solid}. There is no consensus on a single unified definition across the various design and manufacturing applications, which limits the usability or success of CAPP tools that commit to a specific taxonomy. Most process planners are rule-based systems that map individual features (e.g., holes, pockets, slots, and so forth) to specialized manufacturing tools and parameters. These methods are a lot more effective when features are clearly separable, but they rarely extend to complex interacting features \citep{Tseng1998recognition,Tseng1994recognizing}.

\subsection{A Feature-Free Approach}

\begin{figure*}
	\centering
	\includegraphics[width=0.75\textwidth]{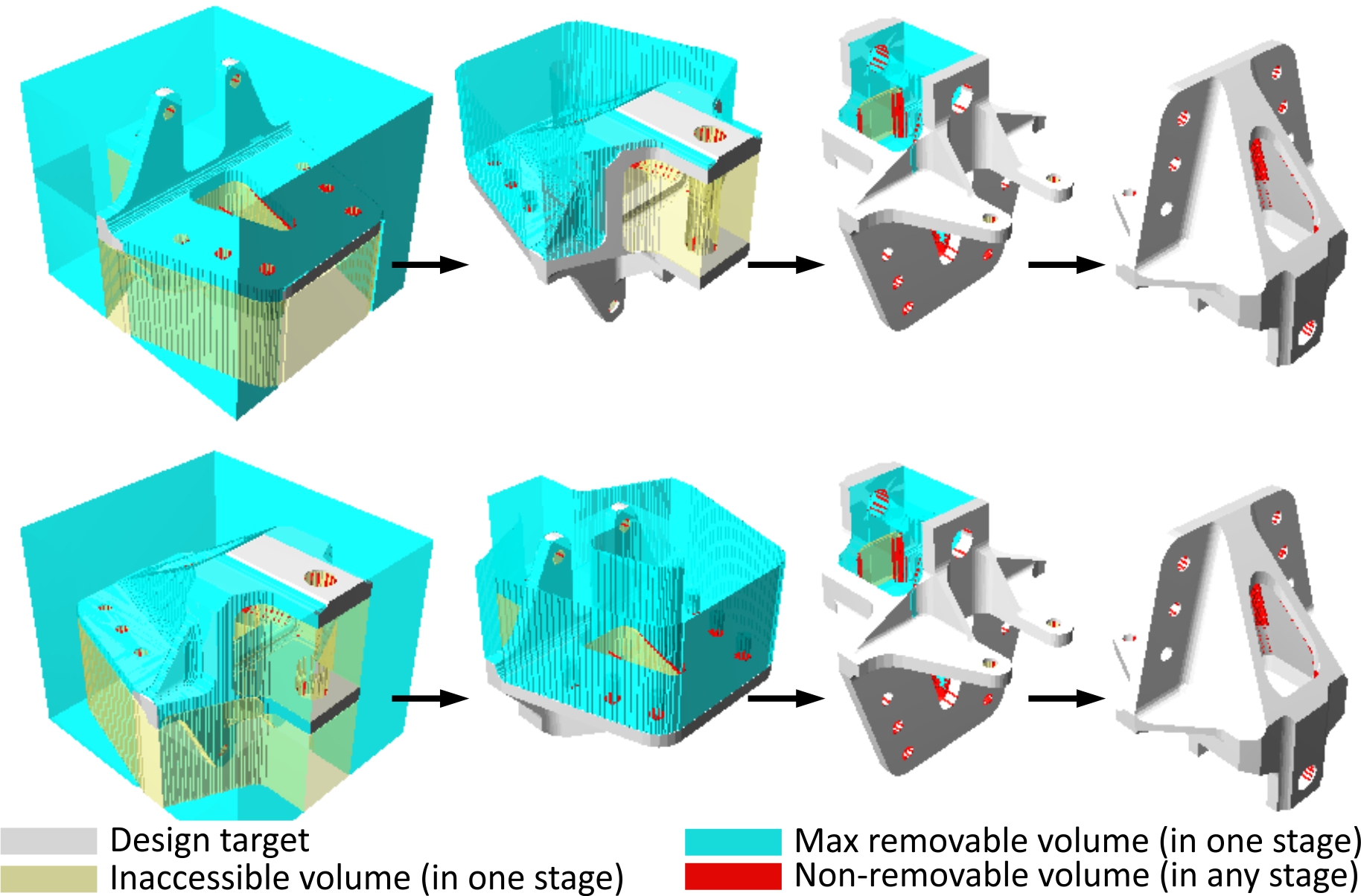}
	\caption{Feature-free process planning for milling \citep{Nelaturi2015automatic}. At every fixturing orientation, the maximal removable volume (MRV) is computed. Different plans (two of them shown here) remove the same MRVs in different fixturing orders. The top plans with the minimum cost are found.}
	\label{fig_smcost}
\end{figure*}

Recently, our group developed a novel feature-free approach to CAPP—with a particular focus on 3-axis milling—based on maximal removable volumes (MRV) at accessible configurations \citep{Nelaturi2015automatic} (Fig. \ref{fig_smcost}). The approach applies to arbitrarily complex part and tool geometries, using group-theoretic configuration space modeling \citep{Lysenko2010group}. The MRVs are computed based on spatial interactions of part and tool in relative motion, conceptualized using morphological operations. This conceptualization facilitates rapid parallel computations by treating 3D shapes as signals and extending basic techniques of digital signal processing to fields in 3D. The morphological operations on sampled/voxelized shapes are mapped to convolutions of their indicator (i.e., characteristic) functions \citep{Serra1983image,Haralick1987image}. Convolutions in the 3D physical domain are, in turn, mapped to pairwise multiplications in the 3D frequency domain via Fourier transforms \citep{Kavraki1995computation}, enabling rapid computations via GPU-accelerated fast Fourier transform (FFT) algorithms. As a result, process plans can be computed using commodity hardware in less than a minute even for arbitrarily complex parts. The algorithm does not rely on semantics of (possibly interacting) features---although having such information can help using specialized tools. This technology has powered a cloud-based quoting platform for machine shops, called \textsf{uFab} (\href{https://ufab.io}{https://ufab.io}).

The underlying theoretical concepts \citep{Lysenko2010group} are general enough to model spatial planning with translations and rotations for high-axis CNC machines. However, there are computational challenges in extending FFT-based convolutions to rapidly generate MRVs with rotational degrees of freedom. In this paper, we extend the earlier implementation for $3-$axis machining to uni-axial turning with translations along or normal to the (unchanging) rotation axis. The goal is to develop the necessary ingredients for combined mill-turn CAPP. However, we do not consider general translations and rotations in this paper.

\subsection{Contributions \& Outline}

\begin{figure*}
	\centering
	\includegraphics[width=\textwidth]{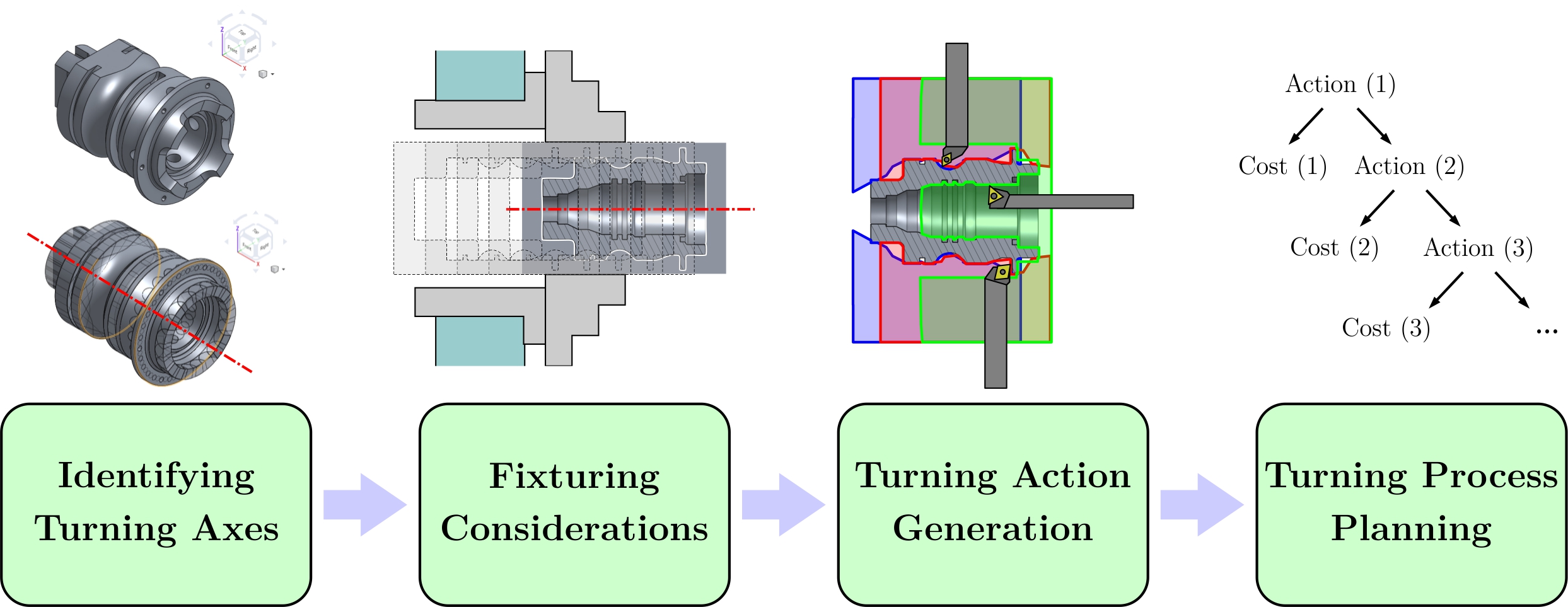}
	\caption{There are 4 main components in our end-to-end process planner for turning: 1) identifying turning axes; 2) fixturing considerations; 3) turning action generation; and 4) turning process planning; discussed in Sections \ref{sec_axis}, \ref{sec_fix}, \ref{sec_act}, and \ref{sec_plan}, respectively. The 3D models in this and following figures are imaginary parts created in \textsf{Onshape} (\href{https://www.onshape.com/}{www.onshape.com}).}
	\label{fig_steps}
\end{figure*}

This paper describes the key ingredients to extend our feature-free approach to turning process planning (alongside milling) with the end goal of enabling mill-turn CAPP. Figure \ref{fig_steps} illustrates the four components of an end-to-end turning process planner, which we elaborate in Section \ref{sec_method}:
\begin{itemize}
	\item Section \ref{sec_axis} presents a generic method to identify one or more candidate axes for turning for a given solid model using the notion of a turnable closure (TC).
	\item Section \ref{sec_fix} discusses fixturing of partially turnable parts via multi-jaw chucks.
	\item Section \ref{sec_act} uses the above results to define a collection of turning ``actions'' based the notion of maximal turnable volumes (MTV).
	\item Section \ref{sec_plan} explains how turnability analysis is performed for predefined sets of turning actions by comparing the complement of the union of all MTVs with the TC, and how the actions are ordered by a planner to minimize the cost.
\end{itemize}
We review some of the earlier work in Section \ref{sec_lit} on volumetric feature-based turnability analysis (Section \ref{sec_lit1}) and turning process planning (Section \ref{sec_lit2}).
%

\section{Related Work} \label{sec_lit}

Over the past few decades, extensive research has been done on feature recognition and process planning for turning. Here we restrict ourselves to a brief overview that is by no means comprehensive.
See \cite{Weill1982survey,Shah1991survey,ElMaraghy1993evolution,Babic2008review,Xu2011computer,AlWswasi2018survey} for thorough reviews of feature-based CAPP methods.

\subsection{Characterizing Turnability} \label{sec_lit1}

\cite{srinivasan1985extraction} and \cite{Li1988part,Li1989part} developed methods for recognition of rotational features (i.e., revolved 2D profiles) and proposed methods to identify non-turning features. The goal was to map CAD data into data structures suitable for integration with CAPP/CAM.
Starting from the observation that most feature identifiers treat rotational and prismatic features differently, \cite{Tseng1994recognizing,Tseng1998recognition} developed a ``machining volume generation'' method to analyze parts with both features. The machining volumes were generated by sweeping part boundaries and classified into machining features/zones by face adjacency relationships.

\cite{Dutta1997feature} also presented a volumetric method for feature extraction for mill-turn machining.
\cite{Yip1997finding} introduced the notions of maximum turnable state (MTS) and maximum turnable volume (MTV), allowing extraction of both isolated and interacting turning features, and introduced algorithms for computing them. Later, \cite{Wilharms1999determination} expanded upon computing and applying the concept of maximum turnability. More recently, \cite{Liu2017extracting} proposed to add additional feature-based semantics to MTS to handle extreme faces or concave interiors.
The MTS and MTV are closely related to our notions of turnable closures and turnable volumes presented in subsequent sections. Importantly, we present a stronger notion of turnability that reaches beyond merely testing for axisymmetry around a given axis and considers the specific (and arbitrary) shape of the available tool inserts, their external/internal accessibility, and collision avoidance between all stationary and moving components on a lathe machine, using configuration space analysis \citep{Lozano-Perez1983spatial}.

Recognizing the difficulties arising from topological alterations caused by feature interaction, \cite{Li2007recognition} argued for the importance of user-defined features incorporated using a neutral feature representation language called ``N-rep.'' Their method extracted the profiles of the revolved faces and detected the remaining non-turnable features. It used a combination of graph-based representation of interactions and rule-based recognition of user-defined features.
In a related work, \cite{Gaines1999custom} presented an extendable representation to allow users to add their custom tool descriptions to the knowledge base of a resource-adaptive feature recognizer.

\subsection{Turning Process Planning} \label{sec_lit2}

The economic impact of optimal parallel mill-turn process planning has been recognized for years. \cite{Levin1996pmps} developed an early prototype CAPP system called \textsf{PMPS} for parallel machining.
\cite{Chiu1999sequencing} formulated the problem as mixed integer programming (MIP) and proposed a genetic algorithm (GA) to optimize operation sequencing.
\cite{Norman2000scheduling} addressed a similar sequencing problem for multiple spindle machines as a scheduling problem with a fixed operation sequence.
\cite{Zhang2003integration} also addressed scheduling optimization and applied iterative modifications to process plans until the predefined optimization objective was obtained.
\cite{Zhang2005integrated} formulated process planning as a nonlinear MIP problem and solved it using a hybrid GA, starting from rule-based heuristics such as min/max processing times.
\cite{Chung2008iso} developed a branch-and-bound based heuristic algorithm to generate nonlinear process plans---i.e., plans that account for alternative manufacturing resources \citep{Kruth1992capp}.
More recently, \cite{mohammadi2012multi} used a hybrid simulated annealing (SA) algorithm for multi-objective optimization of process plans, viewed as a linear MIP problem.
\cite{Su2015advances} considered turning process planning as a multi-objective optimization problem with precedence constraints. They applied a graph-based hybrid GA to avoid infeasible solutions and find process plans with minimal machining cost and maximal utilization.
See the review article by \cite{Shen2006agent} for more details about CAPP research on planning and scheduling. A recent survey of various artificial intelligence (AI) techniques ranging from GA and fuzzy logic to neural networks and knowledge-based systems over the period 1985-2010 can be found in \citep{Deivanathan2019review}.

Our approach to process planning is to map the problem to use combinatorial search techniques such as the A$^\ast$ algorithm and iterative deepening \citep{Russell2016artificial} to find several near-optimal, qualitatively distinct process plans. The results are presented to the user (e.g., manufacturing engineer or machinist) as alternatives, so that they can select the best plan while considering additional criteria that might not have been accounted for by the planner. The planning strategies that we use for turning are similar to the ones used in earlier work on milling \citep{Nelaturi2015automatic}, hence will not be repeated in this paper.

\section{Methodology} \label{sec_method}

Our method does not make any assumptions on the representation or the availability of CAD semantics to facilitate process planning. It can be applied to parts of arbitrarily complex shape in any geometric representation or file format, including polyhedral meshes (e.g., \textsf{STL}) or voxelized models (e.g., \textsf{VDB}) with no feature semantics. However, it does not preclude one from using CAD semantics whenever available; for instance, to look for the rotation axis among the axes of cylindrical and conical surfaces and curves in boundary representations (B-reps) or revolution features in feature-trees to rapidly shortlist the viable options. Below, we described the four components of turning CAPP for arbitrary representation schemes.

\subsection{Identifying Turning Axes} \label{sec_axis}

\begin{figure*}
	\centering
	\includegraphics[width=\textwidth]{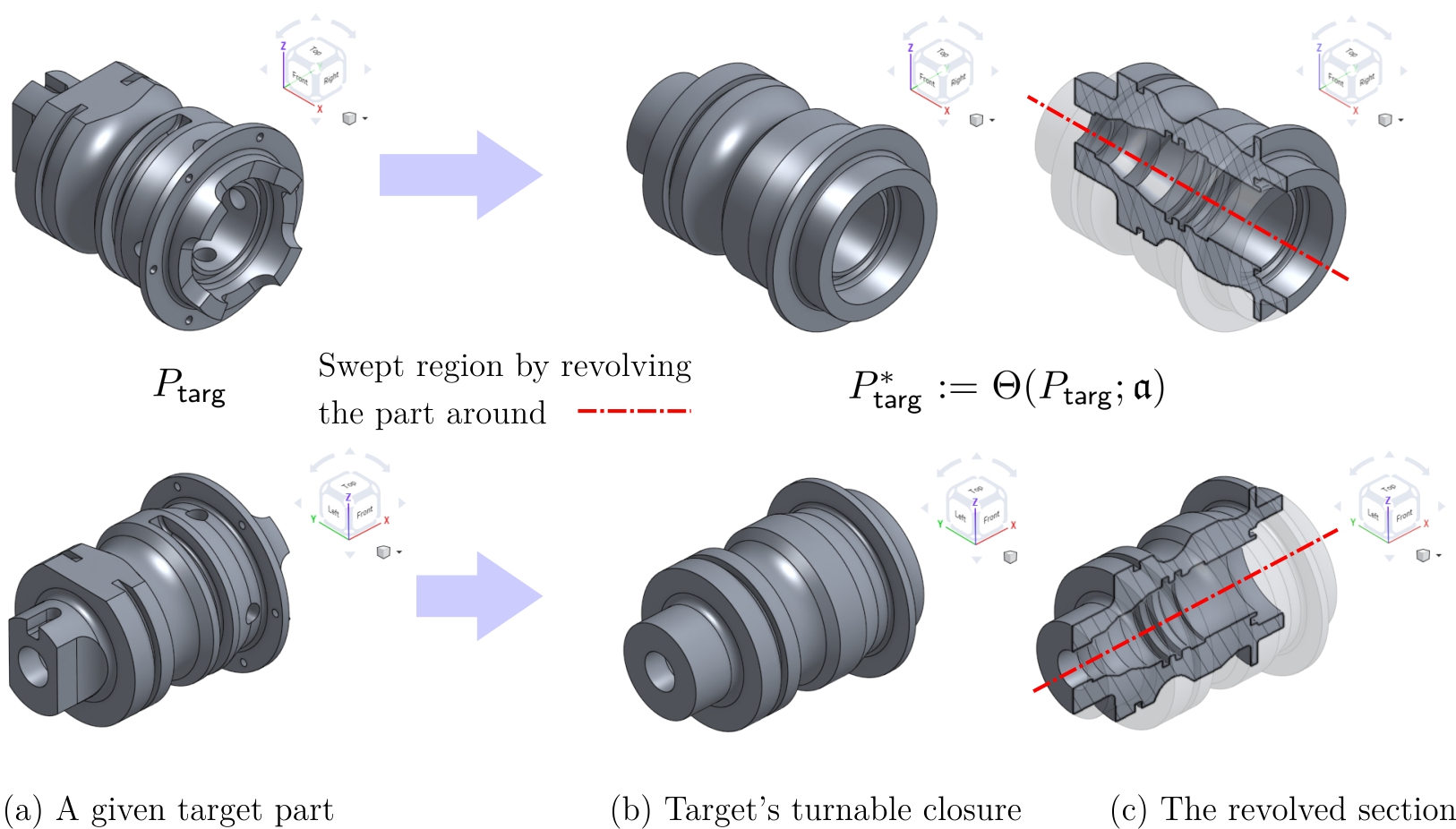}
	\caption{Converting a given target part in (a) to its TC in (b), by sweeping the part for a full rotation around a given axis a (dash-dot red line). The resulting superset (i.e, larger part) can be potentially manufactured by pure turning with the right combinations of tools, and leaves as little volume as possible for other pre-/post-processing machining operations. The same part is shown from two views (top and bottom).}
	\label{fig_closure}
\end{figure*}

For exactly axisymmetric parts (Fig. \ref{fig_closure} (b)) the obvious choice for a turning axis is the axis of rotational symmetry, obtained by solving an eigen-value/eigen-vector problem. Axisymmetric parts have a unique principal axis that coincides with the axis of rotational symmetry, and a class of nonunique pairs of orthogonal principal axes in the plane of cross-section. Although eigen-analysis is a generic and representation-agnostic method of finding the axis, a simpler and faster investigation may suffice when a part---known a priori to be axysimmetric---is given in a representation scheme with explicit information about revolute surfaces (e.g., B-reps or feature-trees).

The challenge is that functional parts are not always axisymmetric (Fig. \ref{fig_closure} (a)), although they might have {\it some} degree of axisymmetry that makes them suitable for turning before/after other (more expensive) machining processes.
Hence, we need a precise measure of {\it approximate} axisymmetry for a part of arbitrary shape around a given axis, which can be determined uniquely and unambiguously regardless of its representation (i.e., data structures) or feature semantics. Our goal is to identify the ``best'' choice of rotation axis when there is no such thing as an axis of symmetry. Eigen-vectors do not necessarily constitute the best choice; however, they do provide a good starting point to gradient-descent search for ``better'' options.

Intuitively, one axis is better than another if it enables producing a solid that is closer to the target by turning alone. In other words, if one removes {\it as much material as possible} from a bar stock that is being turned around a given axis of rotation on a lathe machine, the remaining volume is the closest possible axisymmetric superset of the part. Here, the ``closest'' means the set difference between the turned solid and the target solid has minimal volume among all possible choices of the rotation axis. This would ensure that turning is exploited as much as possible for minimal cost---assuming that turning is one of the cheapest operations and its cost is proportional to the removed volume. The difference can be machined by other means (e.g., milling) in pre-/post-processing. We call this axisymmetric solid the {\it turnable closure} (TC) of the part with respect to a given turning axis.

For a given target part $P_\targ$, we denote its TC with respect to an axis $\axis$ by $P^\ast_\targ := \Theta(P_\targ; \axis)$, where $\Theta(\cdot; \axis)$ is the revolution operator (i.e., sweep for uni-axial rotation) around $\axis$, which is a well-defined solid modeling operator, defined irrespective of the representation scheme \citep{Requicha1980representations}. Note that the TC is a superset of the original part ($P_\targ \subseteq P^\ast_\targ$). One should choose the axis $\axis$ in such a way that the volume of the set difference, denoted by $\vol[P_\targ^\ast - P_\targ]$ is minimized for pre-/post-processing (e.g., milling, drilling, wire-cutting, etc.). This is justified because:
\begin{enumerate}
	\item for a fixed toolset, the machining cost per tooling is approximately proportional to a given tool’s total removed volume; and
	\item turning is often the least expensive of all machining operations, thus it makes sense to leave as little volume as possible for subsequent processes.
\end{enumerate}
The {\it turnability ratio} (TR) of the part is defined by $\gamma(P_\targ) := \vol[P_\targ]/\vol[P_\targ^\ast]$. In general, $0 < \gamma \leq 1$. The closer this ratio is to unity, the more turnable the target part is, leaving a smaller volume fraction $(1 - \gamma)/\gamma$ for the non-turnable features to other machining    processes.

\begin{figure*}
	\centering
	\includegraphics[width=\textwidth]{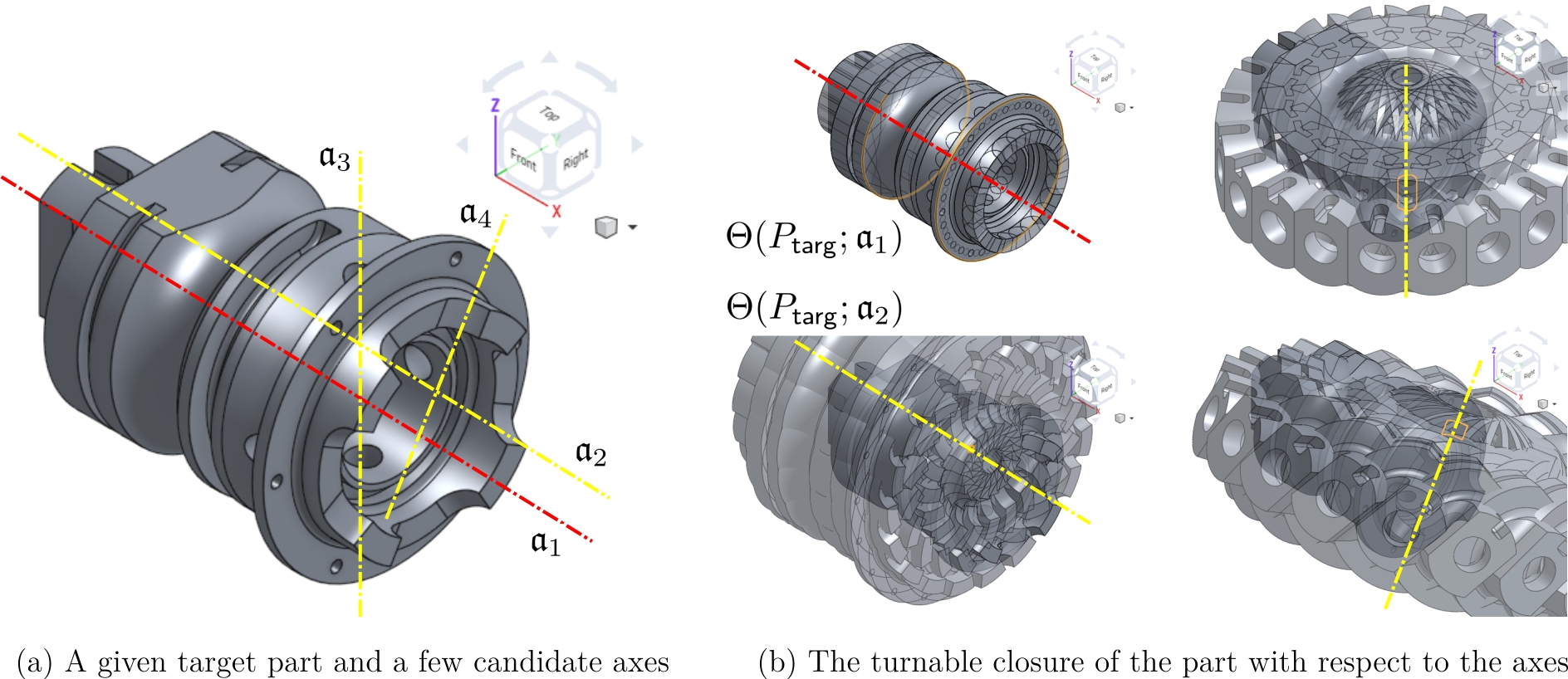}
	\caption{An investigation of the CAD model leads to a set of candidate axes (e.g., of cylindrical faces in this example). One clearly stands out as it results in the smallest TC.}
	\label{fig_axes}
\end{figure*}

\begin{figure*} [ht!]
	\centering
	\includegraphics[width=\textwidth]{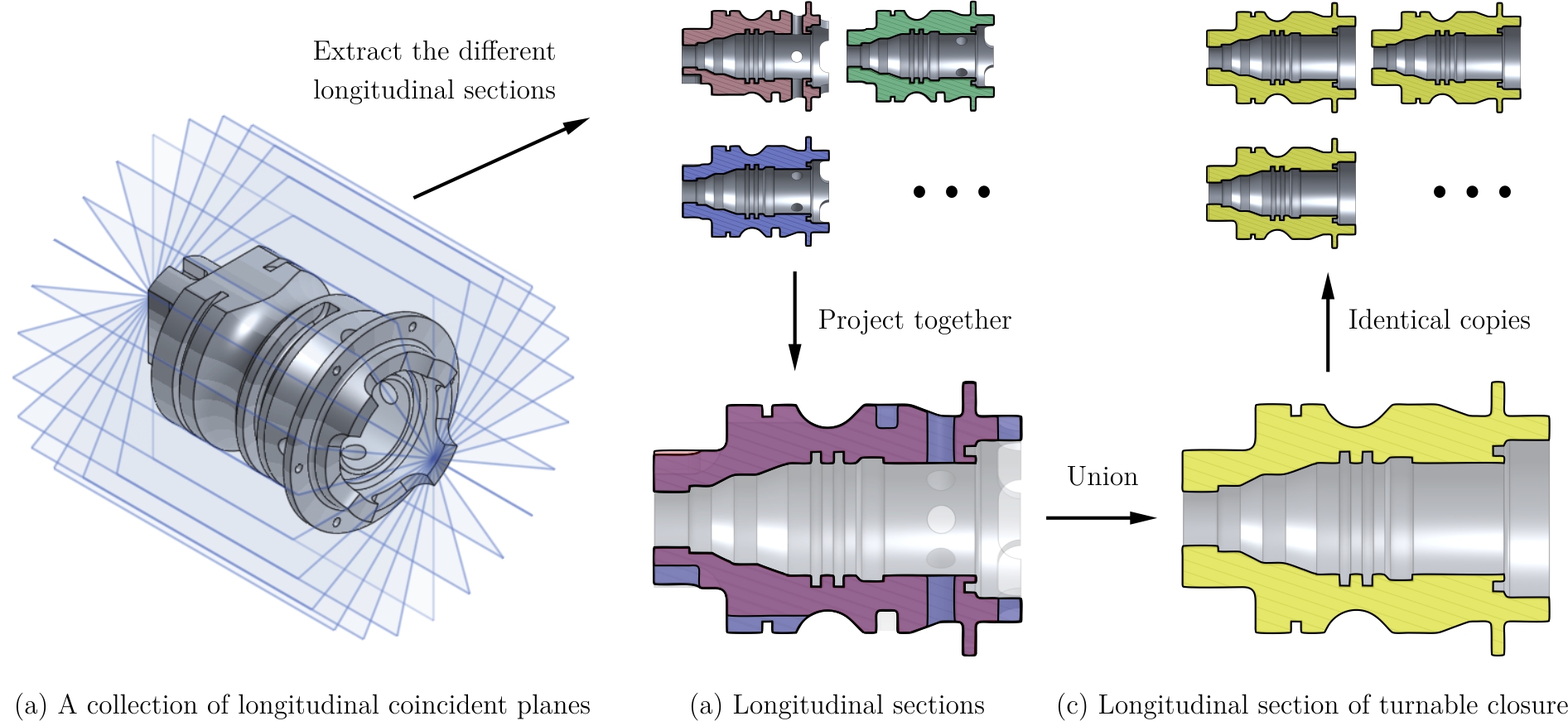}
	\caption{The explicit approach to determining the TC requires projecting a large number of longitudinal sections onto the same 2D plan and computing their union.}
	\label{fig_explicit}
\vspace{1.5em}
	\centering
	\includegraphics[width=\textwidth]{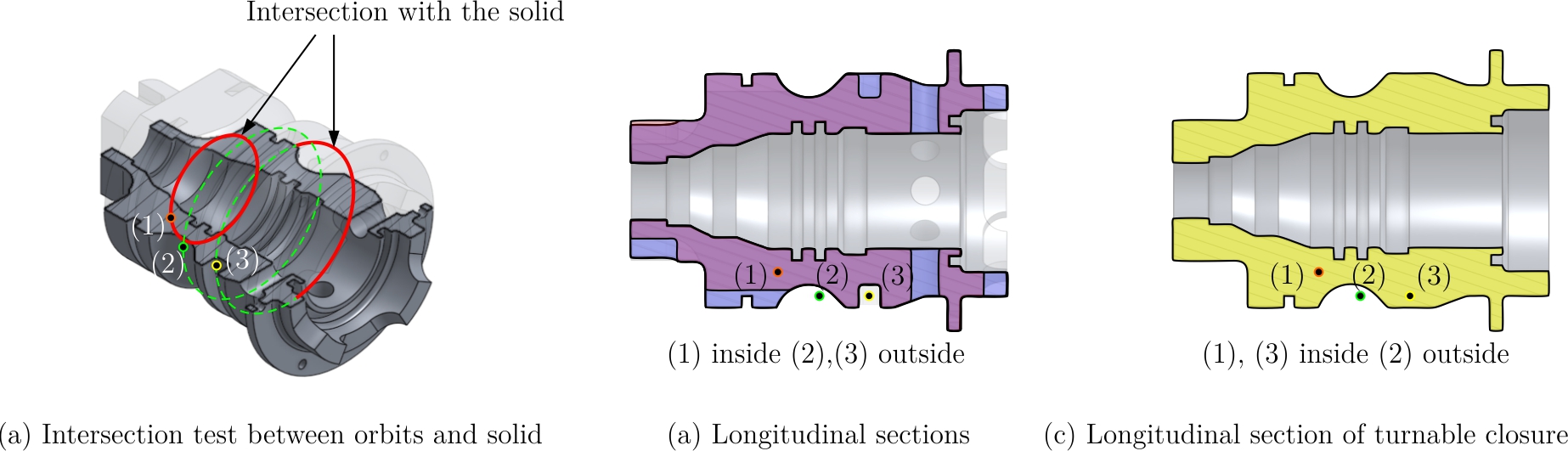}
	\caption{The implicit approach to determining the TC requires a PMC test against it, which is obtained by performing intersection tests between orbits and the solid.}
	\label{fig_implicit}
\end{figure*}

The goal is to find one or more axes $\axis$ that maximize the TR and get it as close as possible to unity. This is an optimization problem that can be solved using any number of gradient-descent techniques. Computing the objective function (i.e., TR) at every iteration requires computing a 3D sweep of the part (Fig. \ref{fig_axes}), which is computationally expensive. However, computing rotational sweeps can be reduced to 2D unions over longitudinal sections or solid-circle intersection tests, which lead to the following dual approaches, respectively:
\begin{itemize}
	\item {\bf Explicit Approach:} The part is sliced into a collection of longitudinal sections. The sections are projected onto the same 2D plane, and their union is computed using planar Boolean operations. These operations can be computed rapidly on polygonal approximations of the sections sliced from polyhedral mesh B-reps (e.g., \textsf{STL}) or on bitmap images sliced from voxel-maps (e.g., \textsf{VTK} or \textsf{VDB}). The TC is obtained explicitly by repeating the same longitudinal section around the given axis of revolution (Fig. \ref{fig_explicit}).
	\item {\bf Implicit Approach:} The TC is never explicitly computed in this approach. Rather, it is implicitly characterized by point membership classification (PMC) devised by intersecting orbits of every query point with the target part. As such, every longitudinal section is rasterized into a 2D image by deciding each point’s in/out classification depending on whether its orbit intersects the target part. This can be interpreted as the 3D voxelization of the axisymmetric TC in a polar-cylindrical frame (Fig. \ref{fig_implicit}).
\end{itemize}
The implicit approach does not rely on computing an explicit representation of the actual revolution volume to obtain the volume fraction, which is all we need to compute the TR for this step. This computation has a perfectly (i.e., embarrassingly) parallel nature, well-suited for a multi-core CPU or many-core GPU.

To find one or more axes that lead to maximum turnability ratios, various sampling techniques can be used. If CAD semantics for the target part $P_\targ$ are available, the CAD model can be mined to find a reasonable collection of axis with high likelihood of being the turning axes (e.g., axes of circular sketches or revolved surfaces). Otherwise, uniform or random sampling can be used followed by standard iterative optimization. As an additional heuristic, the principal axes (of volume or surface) of the target part $P_\targ$ can be used to initially rank the sampled axes based on their closeness to one the principal axis, and shortlist the candidate axes before evaluating their turnability ratios. This is motivated by the fact that for perfectly axisymmetric parts (i.e., $\gamma = 1$) the turning axis is identical with the unique principal axis; therefore, for parts with a high degree of axisymmetry (i.e., $\gamma \approx 1$) the two must be close to each other.

Figures \ref{fig_explicit} and \ref{fig_implicit} illustrate the mechanics of explicit and implicit approaches, respectively, to computing the turnable closures and turnability ratios. The same dual theme will emerge in the subsequent sections for other computations, most notably in Section \ref{sec_act} in which a stronger notion of turnability is presented by considering the precise fixturing constraints, tool geometry, and tool setup.

To simplify the following steps, the world coordinate system (WCS) is chosen to have one of its axes aligned with the spindle's turning axis, and its origin at an arbitrary point on the spindle (e.g., chuck center).
The {\it initial configurations} (denoted by $\map_\init$) are the rigid transformations that map the target part $P_\targ$ and its turnable closure $P_\targ^\ast$ from the arbitrary local coordinate system (i.e., input representation) to the WCS by aligning a given candidate turning axis with the spindle axis. The transformed parts are denoted by $P_\init = \map_\init P_\targ$ and $P_\init^\ast = \map_\init P_\targ^\ast$ as illustrated in Fig. \ref{fig_fix}.

\subsection{Fixturing Considerations} \label{sec_fix}

To simplify the discussion, we restrict our attention to simple two-jaw vises or multi- jaw chucks that require adequate grip at two or more contact points arranged evenly around the spindle. This restriction simplifies automated reasoning for picking the proper grip configuration, i.e., positions and orientations of the workpiece on the lathe machine. Nonetheless, the idea can be generalized to collets, mandrels, face plates, and more complex special-purpose fixtures whose contact geometry requires considering their particular geometric shape and contact properties. See the earlier work by \cite{Nelaturi2015automatic} on automated techniques for modular fixture reconfiguration for milling, using fast hashing algorithms to generate form/force closure grips. These techniques can in principle be extended from milling to turning, but are beyond the scope of this paper.

\begin{figure*}
	\centering
	\includegraphics[width=\textwidth]{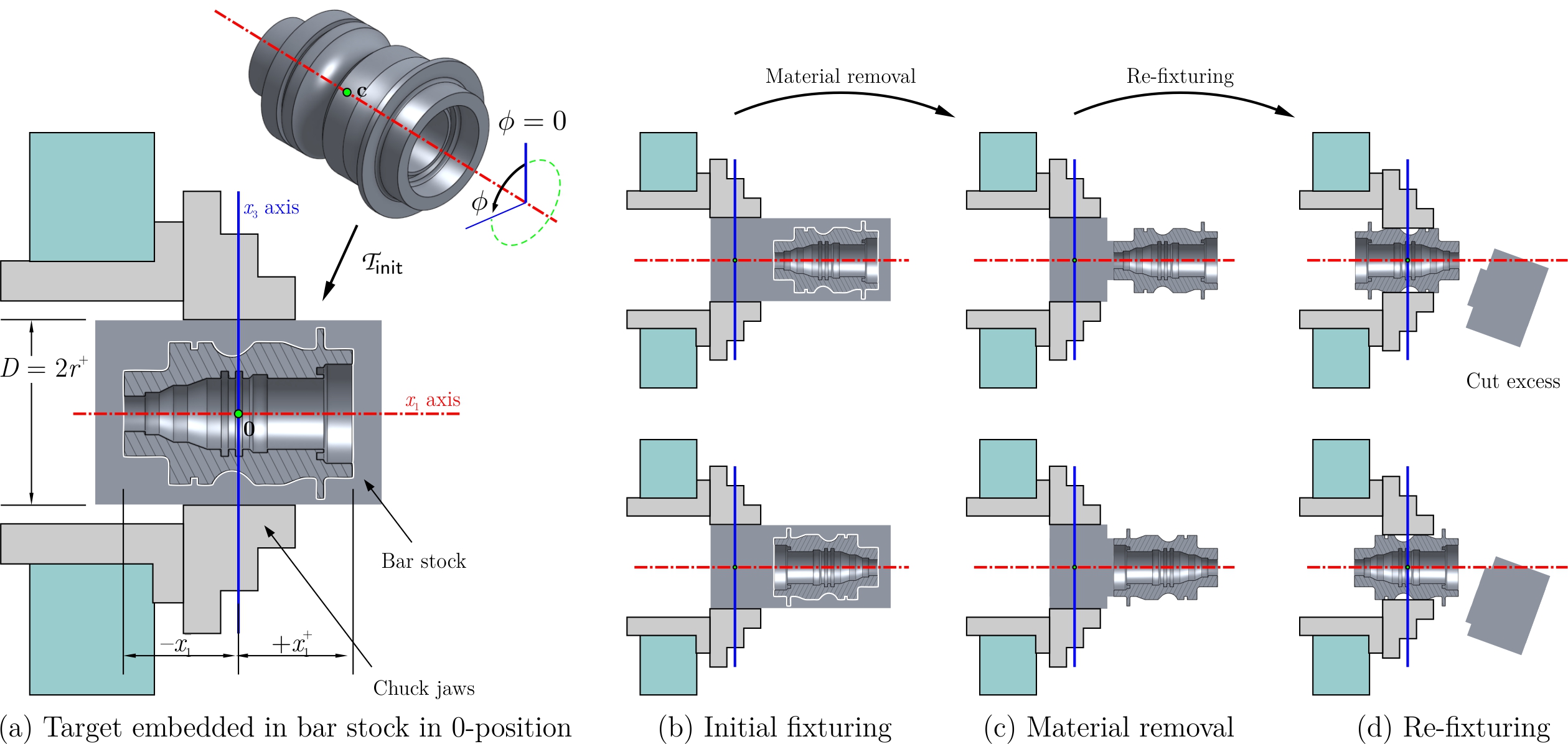}
	\caption{The first transformation positions and orients the part’s TC to align with the machine axis in (a). The different grip configurations are obtained by a discrete set of subsequent rigid transformations (b–d) which are screw motions (top) with or without an additional flip of axis direction (bottom).}
	\label{fig_fix}
\end{figure*}

Figure \ref{fig_fix} illustrates a simple fixturing scenario in which a bar stock is gripped at one of the two ends to turn most of the features without opening the chuck, leaving only the final cutting and finishing to be performed after flipping the workpiece around, re-fixturing, and adjusting the runout with the aid of dial indicators. Every grip configuration has to satisfy an additional condition; namely, at every fraction of a full turn (e.g., one-half, one-third, one-fourth, etc. depending on the chuck), there must exist a cylindrical surface patch of sufficient surface area along the chuck jaws, i.e., a subset of the target part’s boundary has a constant distance from the turning axis and extends along a sufficient length in both longitudinal and circumferential directions.

\begin{figure*} [ht!]
	\centering
	\includegraphics[width=\textwidth]{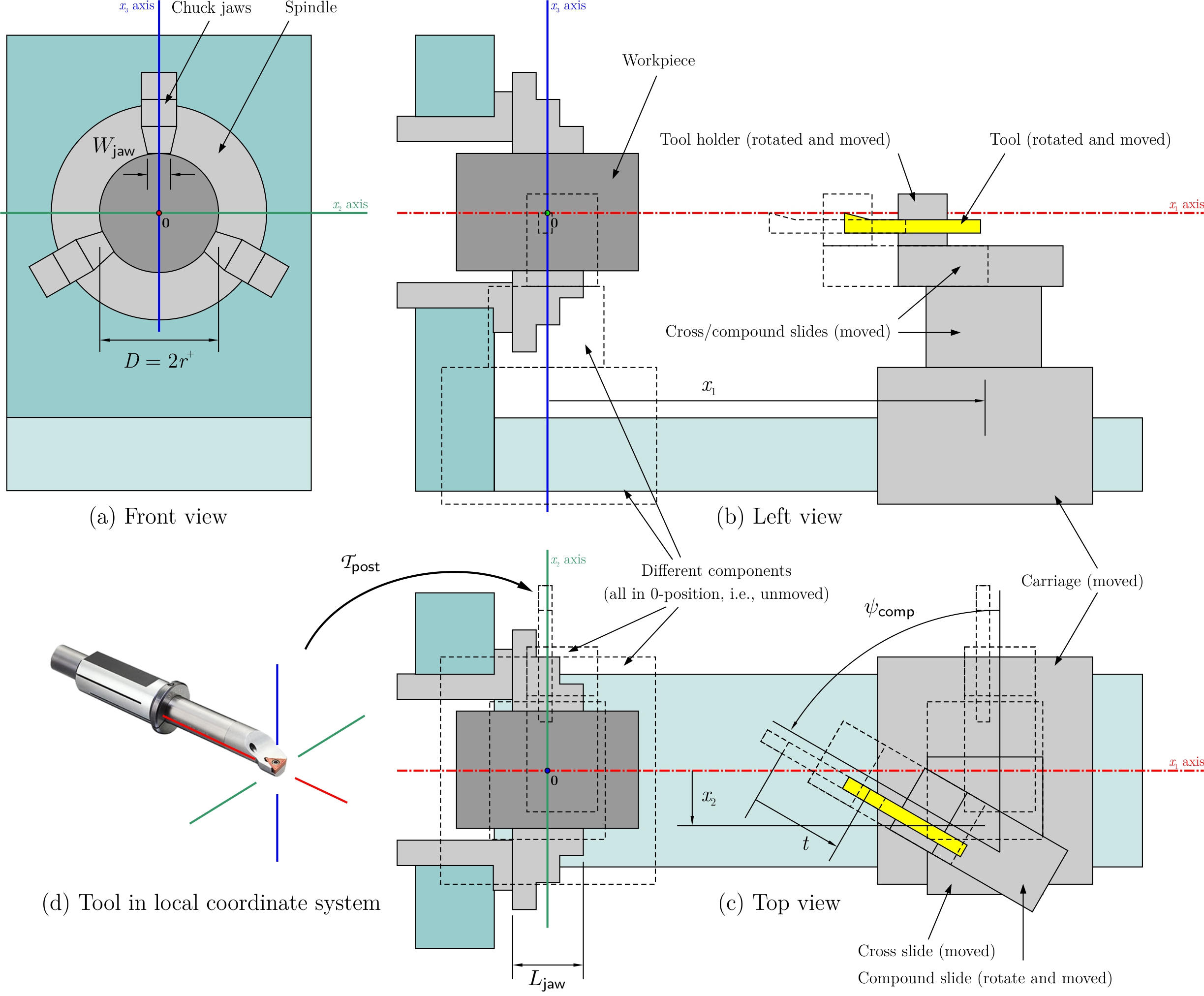}
	\caption{A simplified sketch of the stationary, rotating, and sliding components in the workspace of a simple lathe machine. The planar motion of the sliding components is measure from their 0-position, arbitrarily chosen to be at the world coordinate system’s origin (i.e., spindle center).}
	\label{fig_nom}
\end{figure*}

As illustrated in Fig. \ref{fig_fix} (c), the contact surfaces need not be connected. If we assume the contact between the jaws and the cylindrical surface of the part to happen along line segments, multiple such segments can exist at each jaw. Each segment needs to be lower-bounded in length for proper grip and positioned carefully against the jaw profile to prevent damaging the finished surfaced.
In addition, if the part has pre-existing non-axisymmetric features (e.g., milled in a previous machining step), each jaw may have a different contact profile, all of which must satisfy the proper grip conditions at some initial rotation of the workpiece relative to the spindle.

The {\it grip configurations} (denoted by $\map_\grip$) are the rigid transformations that map the target part $P_\init$ and its TC $P_\init^\ast$ from the initial configuration on the WCS to an adjusted position along and orientation around the spindle axis. The transformed parts are denoted by $P_\fix = \map_\grip P_\init$ and $P_\fix^\ast = \map_\grip P_\init^\ast$.

The {\it fixture configurations} (denoted by $\map_\fix$) are the composite maps $\map_\fix = \map_\grip \map_\init$.
Therefore, we obtain $P_\fix = \map_\fix P_\targ$ and $P_\fix^\ast = \map_\fix P_\targ^\ast$.

\subsection{Turning Action Generation} \label{sec_act}

For every candidate grip configuration along/around the axis, a {\it turning action} is used to as close as possible to the TC with a given tool. Several such actions (with different tools) may be needed before opening the chuck and re-fixturing. Each tool provides a different specialized functionality such as face-, side-, and form-milling, external vs. internal milling, and sawing, different tool insert profiles, hardness properties, surface qualities, and so on.

For a given grip configuration and a particular toolset, the {\it turning actions} are defined in terms of the {\it maximal turnable volumes} (MTV) by each tool at different tool configurations. Each MTV depends on:
\begin{enumerate}
	\item tool insert profile, which determines the smallest removable features; 
	\item tool configuration, which determines reachability (e.g., for internal turning);
	\item collision avoidance between the inactive parts of the moving tool assembly (e.g., tool stem, tool holder, cross/compound slides, and carriage) and the turning chuck/spindle at any relative configuration; and
	\item geometric forming, which requires the to avoid penetrating the target part’s geometry at any relative configuration.
\end{enumerate}
The MTV is the largest subset of the target part’s complement that can be turned by a given tool at a given setting on the tool holder without penetrating into the target shape or collision between stationary, rotating, and sliding machine components.

The first step to formalize the MTV is to consider the degrees of freedom (DOF) of the machine, and all possible modes of relative motion between the part and tool assembly that are constrained due to accessibility and collision avoidance. Figure \ref{fig_nom} shows a schematic of a conventional lathe machine to illustrate its DOF.

We use the common notion of a configuration space obstacle ($\conf-$obstacle for short) \citep{Lozano-Perez1983spatial} to formally define the MTV. The $\conf-$obstacle defines the transformations of the sliding components (i.e., the entire tool assembly) that would result in an undesirable collision event with the stationary and/or rotating parts of the machine (e.g., head and tail stocks, spindle, and chuck’s revolving closure) or with the interior of the target part itself. The undesirable collisions include:
\begin{itemize}
	\item collisions that result in damaging the machine or safety hazards, e.g., the tool assembly collides with the rotating chuck; and
	\item collisions that lead to extra material removal, i.e., the tool insert penetrates into the target shape, cutting more than it should.
\end{itemize}
Note that the tool insert can still interfere with the rotating bar stock to remove the regions {\it outside} the target TC.

Given the TC $P^\ast_\fix$ in its fixtured configuration, let $P^{\ast\ast}_\fix$ denote the union of the TC with the revolution (i.e., rotational sweep) of the chuck $K$ along the turning axis. Recalling from earlier sections that $P^\ast_\fix = \map_\fix \Theta (P_\targ; \axis)$, we can account for the chuck by letting $P^{\ast\ast}_\fix = \map_\fix \Theta (P_\targ \cup K; \axis)$, where $\map_\fix = \map_\grip \map_\init$. Consider a tool assembly $T = (H \cup C)$ where $C$ is the cutter (i.e., tool insert) and $H$ is the remaining moving parts. The $\conf-$obstacle is defined by $\obs(P^{\ast\ast}_\fix, T)$.

Let us assume that the tool orientation remains fixed during every turning, i.e., every time the tool is re-oriented on the tool holder, whatever it removes afterwards counts as a new action. We use a fundamental result in spatial reasoning that the $\conf-$obstacle for translations can be computed as the (interior of) Minkowski sum of the stationary and reflection of moving components \citep{Lysenko2010group}:
\begin{equation}
	\obs(P^{\ast\ast}_\fix, T) = (P^{\ast\ast}_\fix \oplus T^{-1}) = \big(\map_\grip \map_\init \Theta (P_\targ \cup K; \axis) \big) \oplus (H \cup C)^{-1}
\end{equation}
where $(\cdot)^{-1}$ stands for reflection with respect to the WCS origin.
If $W$ represents the set of all translations accessible to the tool, bounded by the motion range of the carriage and cross/compound slides, the maximal set of translations the result in no undesirable collisions is $\free(P^\ast_\fix, T) := W - \obs(P^\ast_\fix, T) = (P^{\ast\ast}_\fix \oplus T^{-1})^c$, where $(\cdot)^c$ stands for set complement (i.e., ``negative'' space) in $W$.

The 3D sweep of the tool insert along this maximal set of motions yields the maximal volume in the 3D space that does not result in undesirable collisions. This is obtained as a Minkowski sum of the maximal set of motions with the tool insert \citep{Nelaturi2015automatic}, followed by an additional revolution (i.e., rotational sweep). The latter converts the maximal volume swept by the tool to the maximal volume removed from the bar stock due to rapid turning:
\begin{equation}
	Q(P^{\ast\ast}_\fix, H, C) = \Theta \big( \free(P^{\ast\ast}_\fix, T) \oplus C; \axis \big) = \Theta \big( (P^{\ast\ast}_\fix \oplus T^{-1})^c \oplus C; \axis \big)
\end{equation}
The tool geometry is commonly such that the MTV can be computed by sweeping only the largest horizontal cross-section (i.e., top face) of the tool insert in two steps: first along the maximal accessible configurations (i.e., 2D translations); then around the turning axis. With these assumptions, the problem reduces to computing Minkowski operations in 2D, between longitudinal section of the target part’s TC and the tool insert cross-section. Once again, the two alternative approaches are:
\begin{itemize}
	\item {\bf Explicit Approach:} The longitudinal section of the target part’s TC in its fixtured configuration obtained earlier is used to compute the longitudinal section of the MTV by Minkowski operations in 2D. These operations can be implemented rapidly on polygonal approximations of the sections sliced from polyhedral mesh B-reps (e.g., \textsf{STL}) or on bitmap images sliced from voxel-reps (e.g., \textsf{VTK} or \textsf{VDB}). The MTV is obtained explicitly by repeating the longitudinal section around the given axis of revolution.
	\item {\bf Implicit Approach:} Earlier, we devised a PMC test for the target part’s TC in its fixtured configuration, which led to a 2D image representation of its longitudinal section. The PMC test against the MTV can be conceptualized as a {\it convolution} applied to this image \citep{Kavraki1995computation}, using the tool’s cross-section image as a filter. The resulting 2D image is the longitudinal section of the MTV. This can be interpreted as the 3D voxelization of the axisymmetric MTV in a polar-cylindrical frame.
\end{itemize}
Once again, the implicit approach has several computational advantages. Most importantly,
convolutions on 3D images can be computed rapidly via FFTs \citep{Kavraki1995computation} for which optimal parallel implementations exist on both CPU and GPU.

\begin{figure*} 
	\centering
	\includegraphics[width=\textwidth]{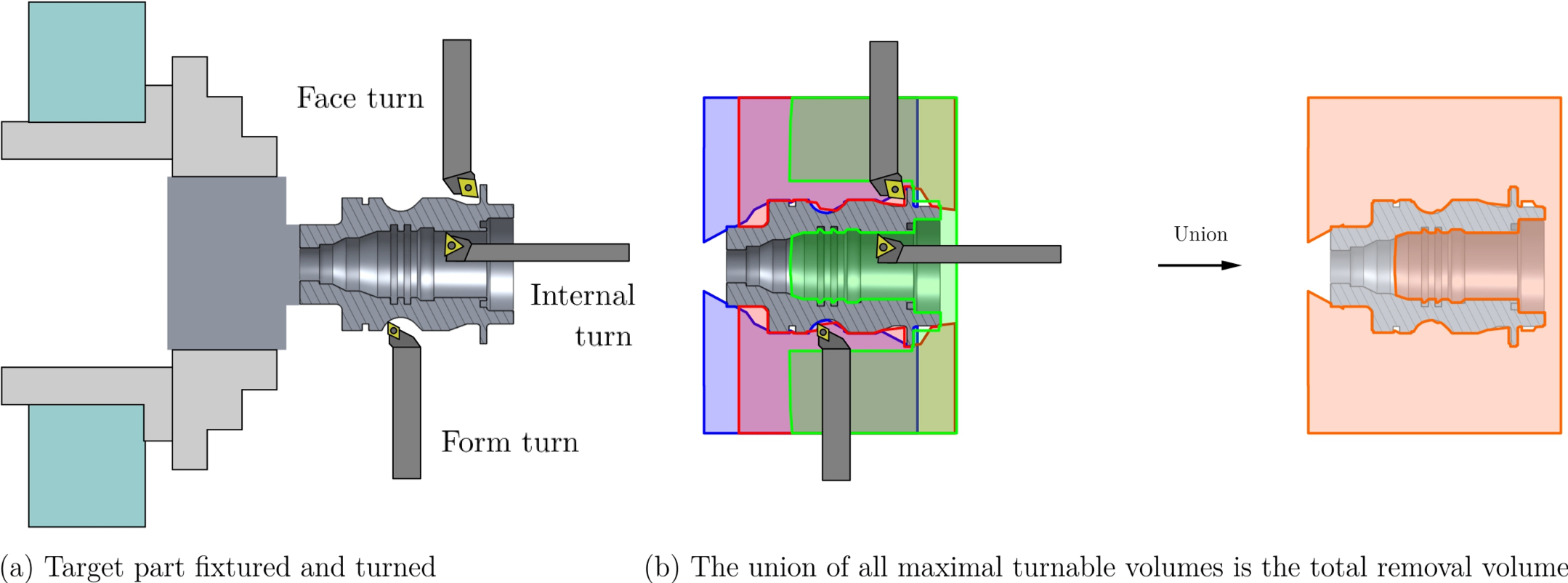}
	\caption{The MTVs for different tool profiles and setup configurations with the same grip/fixturing configuration. The computations are reduced to Minkowski operations on the longitudinal sections.}
	\label{fig_test}
\end{figure*}

It is also worthwhile emphasizing that the commonly used notions of ``accessibility'' or ``reachability'' are implicitly encoded into the MTVs, whose rapid computation does not rely on any simplifying assumption on the part or tool geometry, or that of any other component in the lathe machine’s workspace.

Figure \ref{fig_test} illustrates the MTVs for a few tools at different tool setup. The same tool can generate different MTVs for each choice of turning axis, grip configuration, and tool setup configuration. Each choice and its corresponding MTV define a separate turning “action” generated independently (and in parallel).

\subsection{Turning Process Planning} \label{sec_plan}

Once a collection of actions and their effects in terms of the MTVs are precomputed, the combinatorial space of all possible sequences of actions (i.e., process plans) is a partially ordered set of state transitions for the workpiece as it evolves from being a bar stock to the complement of the union of all MTVs. Every action’s cost is assumed to be proportional to the total volume removed, which is computed as the intersection of that action’s MTV with the action’s input (i.e., intermediate state) at each step. 

Importantly, an early manufacturability test can be defined by simply computing the union of all MTVs for any subset of actions, removing the union from the bar stock, and comparing it with the part’s TC (Fig. \ref{fig_test}). Given several tools $T_1 = (H \cup C_1)$, $T_2 = (H \cup C_2)$, $\ldots$, $T_n = (H \cup C_n)$, each represented in the properly oriented cofiguration on the tool holder, the as-turned shape is obtained by:
\begin{equation}
	P^\dag_\fix = S - \Big( Q(P^{\ast\ast}_\fix; H, C_1) \cup Q(P^{\ast\ast}_\fix; H, C_2) \cup \cdots \cup Q(P^{\ast\ast}_\fix; H, C_n) \Big)
\end{equation}
i.e., $P^\dag_\fix = S - \bigcup_{1 \leq i < n} Q_i$,
where $S$ stands for the bar stock. In general, the as-turned shape is a superset of the target shape ($P^\ast_\fix \subseteq P^\dag_\fix$) due to small non-removable features (e.g., due to high-curvature fillets) or inaccessible tool configurations (e.g., due to tool stem and part collision). The {\it turnability test} for the given set of tools is to check whether the set difference $(P^\dag_\fix - P^\ast_\fix)$ is small enough (e.g., within tolerance). This is a stronger test for turnability than axisymmetry used to obtain the TC, taking into account the precise shape and orientation of the tool assembly, the smallest features it can remove, and the regions it can reach in the part. If the test fails, the part is not manufacturable with the given set of actions created from the sampled
turning axes, grip configurations, setup configurations, and tool insert
profiles. For turnability test, the union operation can be done in any order, regardless of the actual order in which the actions are called in a given process plan. Hence, the manufacturability test can be done prior to process planning.

\begin{figure*} 
	\centering
	\includegraphics[width=\textwidth]{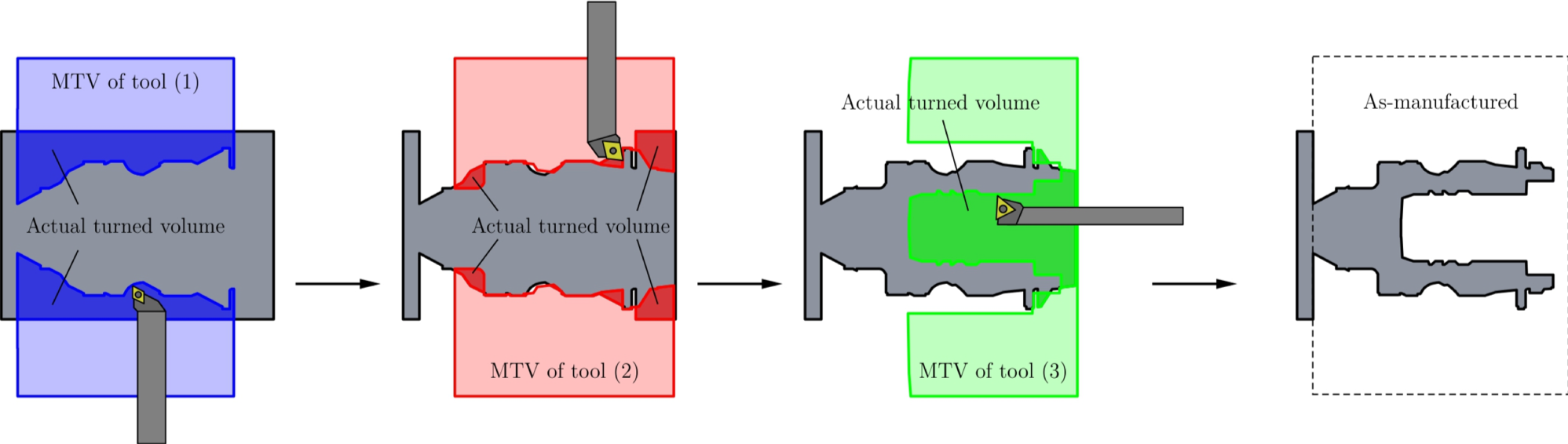}
	\caption{An example turning process plan. The intersection of the MTV with the workpiece at each intermediate state is the actual turned volume, which is used to estimate the cost of that step.}
	\label{fig_mtv}
\end{figure*}

If the TC passes the test, a standard AI search algorithm (e.g., the A* algorithm) \citep{Russell2016artificial} can be used to find cost-optimal plan(s) (Fig. \ref{fig_mtv}). The cost is typically defined in terms of the lathe hour-rate, modified to account for the tool wear-and-tear (e.g., cost per time of operation) denoted by $c_i > 0$  and tool feed rate (e.g., volume removal per time) denoted by $f_i > 0$. If the $i-$th action removes a volume of $v_i \geq 0$, the expected cost of the action is $\cost[v_i] = c_i v_i /f_i$. The removed volume can be obtained as the volume of the set difference between action's MTV $Q_i = Q(P^{\ast\ast}_\init; H, C_i)$ and the workpiece in its intermediate state, which depends on the previous actions. If a plan is specified by the permutation of the indices such that $i= 1, 2, \ldots, n$ represent the order in which the actions are applied, the total cost is:
\begin{equation}
	\cost = \sum_{1 \leq i \leq n} \cost[v_i] = \sum_{1 \leq i \leq n} \frac{c_i}{f_i} \vol \Big[ Q_i - \big(  S - \bigcup_{1 \leq j < i} Q_j \big) \Big]
\end{equation}
Noting that $S - \bigcup_{1 \leq j < i} Q_j$ represents the workpiece in its intermediate state after applying the first $(i-1)$ actions, i.e., removing the first  $(i-1)$ MTVs from the bar stock. Figure \ref{fig_mtv} illustrates one possible plan and the longitudinal cross-section of the actual removed volumes for the MTVs shown in Fig. \ref{fig_test}.

\section{Conclusion}

With the wide availability of high-axis CNC machines with mill-turn functionalities, it is paramount for a practical CAPP solution to be able to reason about both milling and turning for complex designs with interacting features. Parts produced via turning are generally cheaper due to higher availability of lathe machines and lower machine-hour rates and tool inserts compared to milling. In addition, the surface quality and precision obtained by turning is usually unattainable by other means, even if machining the nominal geometry is feasible (e.g., via milling).

In this paper, we reported on extending a powerful feature-free approach for CAPP \citep{Nelaturi2015automatic} based on configuration space modeling \citep{Lozano-Perez1983spatial} from milling to turning. For parts of arbitrarily complex shape---with possibly missing or complicated feature semantics---our framework is capable of automatically determining manufacturability for a given set of tools and grip/setup configurations in terms of maximal removable volumes. It further generates a sequence of high-level turning actions by rapidly searching through the space of possible process plans for the most cost-effective solutions.

\paragraph*{Acknowledgements.}

This research was funded by Sandvik Coromant. The responsibility for any errors or omissions lies solely with the authors.

\bibliographystyle{plainnat}
\bibliography{turnPlanning}

\end{document}